\documentclass[twocolumn,prl]{revtex4}
\usepackage{graphicx}
\usepackage{amsmath}
\usepackage{amssymb}

\begin{document}

\title{Detecting measurement-induced relative-position localisation}
\author{P.A. Knott, J. Sindt, and J.A. Dunningham}
\address{School of Physics and Astronomy, University of Leeds, Leeds LS2 9JT, United Kingdom}

\begin{abstract}
One interpretation of how the classical world emerges from an underlying quantum reality involves the build-up of certain robust entanglements between particles due to scattering events \cite{Rau2003a}. This is an appealing view because it unifies two apparently disparate theories. It says that the uniquely quantum effect of entanglement is associated with classical behaviour. This is distinct from other interpretations that says classicality arises when quantum correlations are lost or neglected in measurements. To date the weakness of this interpretation has been the lack of a clear experimental signature that allows it to be tested. Here we provide a simple experimentally accessible scheme that enables just that. We also discuss a Bayesian technique that could, in principle, allow experiments to confirm the theory to any desired degree of accuracy and we present precision requirements that are achievable with current experiments. Finally, we extend the scheme from its initial one dimensional proof of principle to the more real world scenario of three dimensional localisation.

\end{abstract}
\maketitle


\section{Introduction}

The boundary between quantum and classical physics has long been a perplexing issue for physicists. Why should one set of rules apply to one size scale and another set of rules apparently apply to another? More vexing perhaps is the fact that the boundary that distinguishes the two sets of rules is not sharp. So it is not always clear which theory should be applied.

A lot of work has been done to understand this boundary and the prevailing view is that it can be interpreted in terms of decoherence \cite{Joos1985a,Ghirardi1986a,Zurek1991a,Zurek2003a}. Simply put, this says that quantum systems tend to interact with their environments and become entangled with them. The total system including the environment is therefore properly treated with quantum physics. However, if we are interested only in the quantum subsystem and make measurements only on this, we effectively throw away the information about which environmental states are correlated with which subsystem states. We then find that the subsystem appears to behave more and more classically the more it has interacted with the environment. In effect, by throwing away information about the quantum correlations we are left with a system that behaves classically. 

Another interpretation that extends this idea was put forward a few years ago \cite{Rau2003a}. It showed the emergence of classicality without having to throw away all the information about the quantum correlations. In fact it showed that classicality can be related to quantum entanglement between different subsystems. In this interpretation, even classical objects are entangled with one another but with a special type of robust entanglement sometimes called ``fluffy-bunny" entanglement \cite{Wiseman2004a, Dunningham2005a}. This is an appealing view since it gives one consistent theory that describes both quantum and classical systems.
It is also intriguing that in this theory classicality is associated with entanglement, which is usually thought of as a purely quantum feature. Furthermore, this interpretation gives a clear basis to the idea that we should only think in terms of relative (rather than absolute) positions in physics.

Despite all the pleasing features of this formalism, up until now it has suffered from the major flaw that it has not been at all clear how it could be tested experimentally. In this paper we resolve this issue by providing a simple, experimentally accessible scheme that could clearly demonstrate this process. This brings the idea into the realm of a testable physical theory. We begin by reviewing the scheme \cite{Rau2003a} for measurement-induced relative-position localisation through entanglement. We then describe the experimental scheme that presents a signature for detecting this process. We finish by showing how the localisation can be extended to particles in three dimensions.

\begin{figure}
\centering
\includegraphics[scale=0.4]{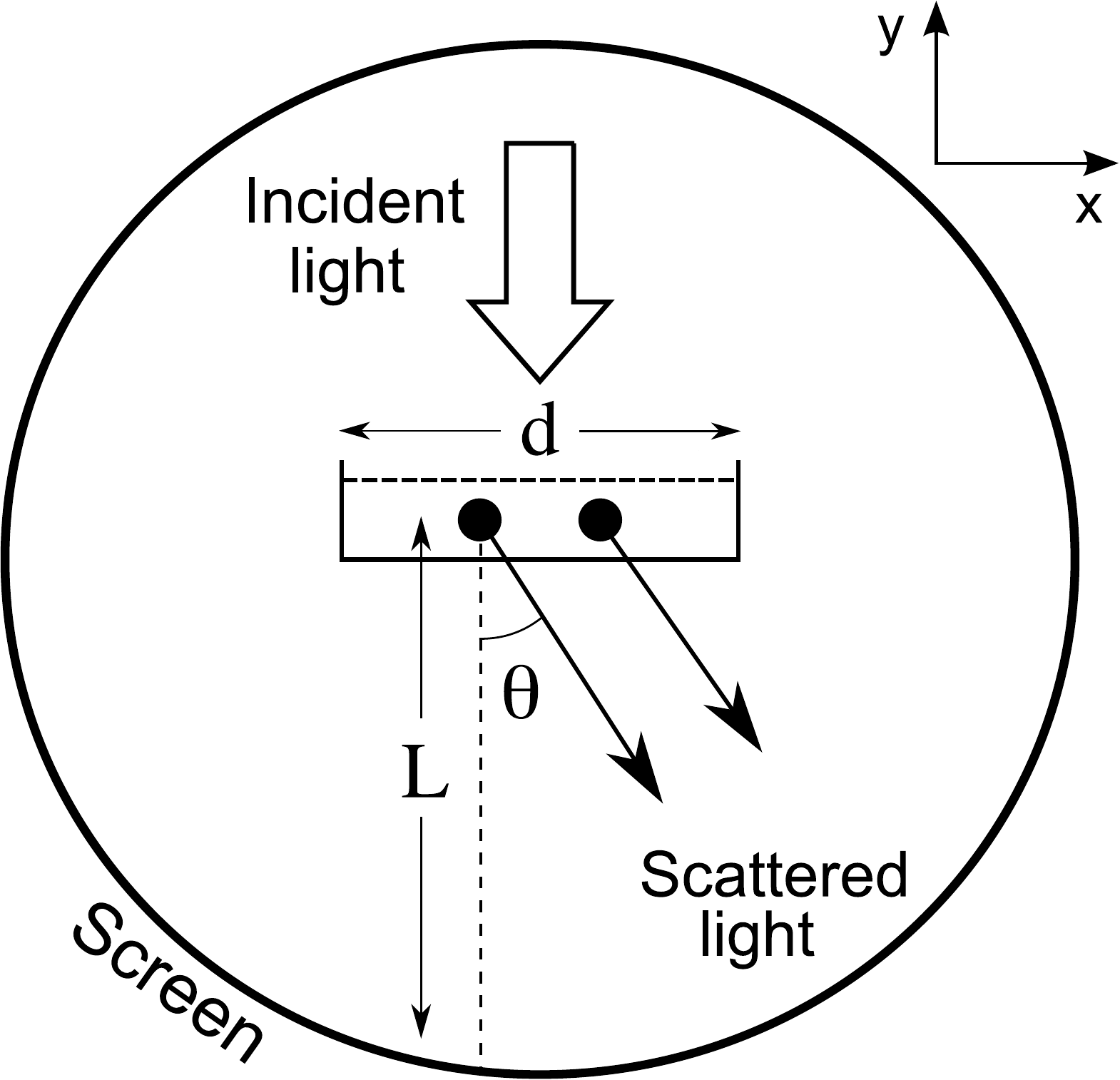}
\caption{A schematic showing the setup. Two massive particles are delocalised over some region $d$ and are illuminated by plane wave incident light. The scattered light is detected at an angle $\theta$ on a screen located at a distance $L$ away from the particles. For clarity, the diagram is not to scale: we consider the case where $L\gg d$.}
\label{fig:schematic}
\end{figure}


\section{Localisation in One Dimension }

The setup is shown in Fig.~\ref{fig:schematic}. Two particles are delocalised over some region $d$ in the $x$-direction in the sense that their de Broglie wavelengths are comparable to $d$ in this dimension. We consider the case of distinguishable massive non-interacting particles that are tightly confined in the $y$ and $z$ directions. These two particles will form the sub-systems of the system we are interested in. They are illuminated with plane-wave light with wavelength $\lambda$ incident along the $y$-axis, which scatters from them and is detected at an angle $\theta$ on a screen located at a distance $L$ away. We will consider the far-field limit where $L\gg d$. The photons are the environmental states of our scheme.

The initial wave function of the particles is $c(x)$ where $x$ represents the \emph{relative} position of the two particles. Since the particles are delocalised over $d$, the position of one particle relative to the other lies in the range $[-d,d]$. Strictly, to give a full spatial description of the system we should specify the centre-of-mass position as well as the relative position. However the centre-of-mass remains unentangled from the relative position coordinate throughout the scattering process and so can be conveniently neglected \cite{Rau2003a}.
When a photon of wavelength $\lambda$ scatters off a particle into angle $\theta$, the particle receives a momentum kick in the $x$-direction of $\Delta p = h\sin\theta/\lambda$, where $h$ is Planck's constant. In relative momentum space the particles therefore receive a kick of $\pm h\sin\theta/\lambda$ depending on which particle the photon scatters from and, since we do not know, we get a superposition of both possibilities.
This allows us to write the overall wavefunction of the system after a photon has scattered as:

\begin{eqnarray}   
\label{scatterwf}
\Psi(x,\theta) = 
\begin{cases}
\frac{1}{2\sqrt{2\pi}}c(x) \left( e^{\frac{i2\pi x}{\lambda}\sin\theta} + e^{\frac{-i2\pi x}{\lambda}\sin\theta} \right) & \text{if } \theta \not= 0 \label{label1} \\ \\
c(x)A(x) & \text{if } \theta=0
\end{cases}
\end{eqnarray}
The term defined as:
\begin{eqnarray}
A(x) = \left[\frac{1}{2\pi}\int_{0}^{2\pi} \sin^2 \left( \frac{2\pi x}{\lambda}\sin\theta' \right) \, d\theta'\right]^{1/2},
\end{eqnarray}
represents a nonscattering event that leaves the photon in the undeflected state. This term is necessary because the total rate of scattering (integrated over all angles) depends on the separation of the particles, $x$. Odd as it seems at first sight, this means that detecting a photon that is not scattered gives us information about the relative position of the particles. The $A(x)$ term is required to properly account for this.

\begin{figure}[b]
\centering
\includegraphics[scale=0.47]{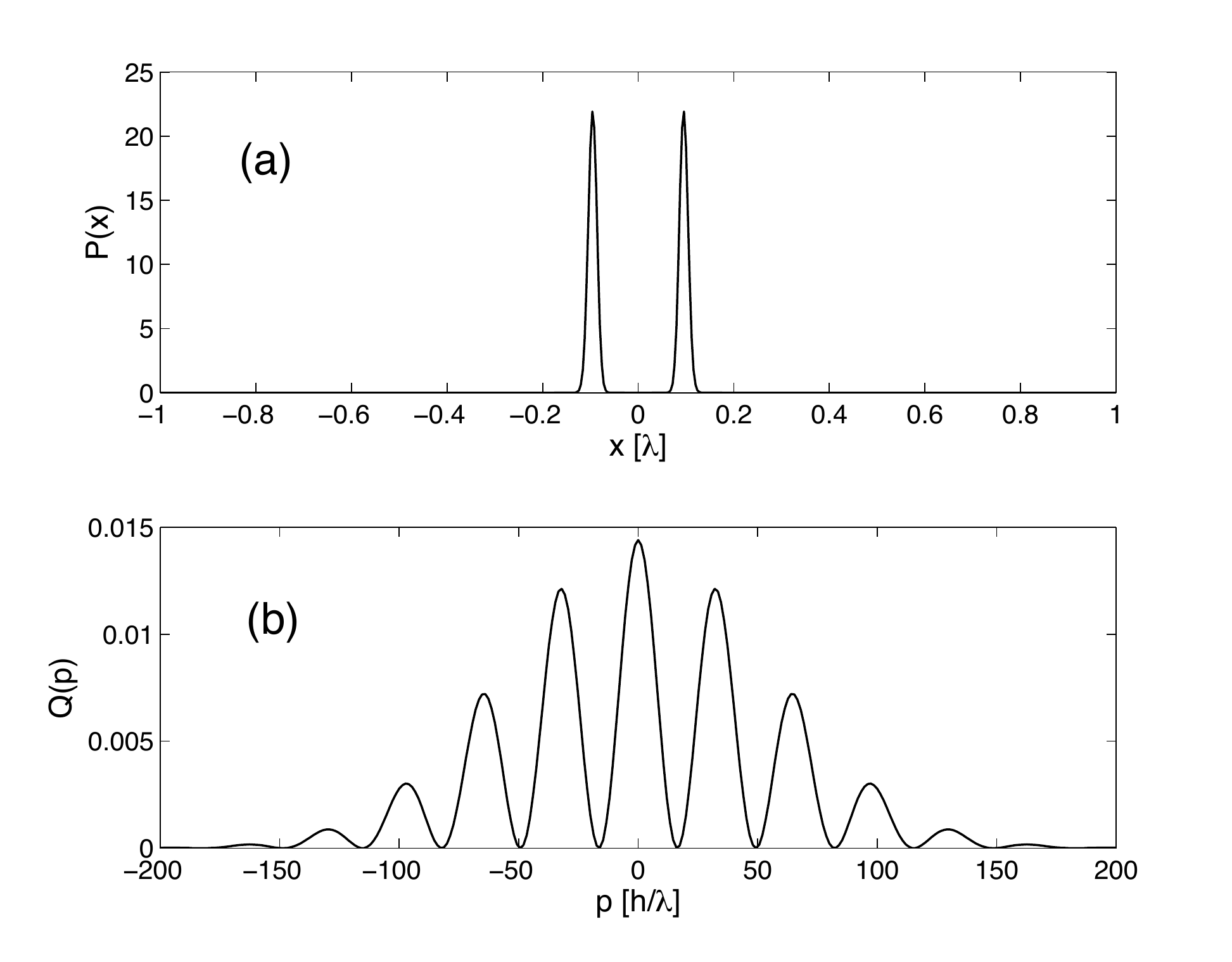}
\caption{The case of light scattering causing relative localisation. (a) Probability density, $P(x)$, for the relative position of the particles after the scattering and detection of 150 photons. The position is given in units of the wavelength, $\lambda$, of the scattered light. (b) Probability density, $Q(p)$ for the corresponding relative momentum of the particles.}
\label{fig:localise}
\end{figure}

The probability density for detecting a scattered photon at angle $\theta\neq 0$, and the probability density of detecting a nonscattered photon, $\theta=0$, are given by: 
\begin{align*}
&P_S(\theta) = \frac{1}{2\pi}\int_{-d}^{d}|c(x)|^2
\cos^2 \left( \frac{2\pi x}{\lambda}\sin\theta \right) \, dx \\
&P_{NS} = \int_{-d}^{d} |c(x)|^2 A^2\, dx = 1 - \int_{0}^{2\pi}P_{s}(\theta)\, d\theta.
\end{align*}
To model a scattering event we generate a random number to see whether the photon is scattered and, if so, at what angle. If it is not scattered the (unnormalised) new state is $c(x)A(x)$ and if it is scattered at an angle $\theta_1$, the (unnormalised) new state is given by:

\begin{eqnarray*}
\psi(x,\theta_1) = c(x)\cos \left( \frac{2\pi x}{\lambda}\sin\theta_1 \right).
\end{eqnarray*}
We then normalise the state and repeat for the next photon.

We choose to start our simulations with a flat distribution, $c(x) = 1/\sqrt{2d}$ because we want to choose the `hardest' case and show that relative localisation builds up even when there is none to begin with. This choice does not  restrict the generality of the results and qualitatively similar outcomes  are obtained for different choices. The probability distribution, $P(x)$, for the relative position of the two particles is shown in Fig.~\ref{fig:localise}(a) for a typical run after 150 photons have been detected and for $d=\lambda$. We assume that the 150 photons are all incident on the particles in a sufficiently short time period that we do not need to consider the dynamics of the particles between detection events. Initially the distribution is completely flat and we see that the measurement process has induced localisation. We have checked the variance of each of the peaks in Fig.~\ref{fig:localise}(a) and have shown that it varies inversely with the number of scattered photons.

The above analysis is for monochromatic light of wavelength $\lambda$. We have also  simulated the localisation process for ambient light. For each scattering event the wavelength of the photon is selected from a blackbody distribution. We have used several different temperature blackbody distributions, and we find that in all cases the relative position localisation takes place.


\section{Proposed Experiment to Test the Localisation}

\begin{figure}
\centering
\includegraphics[scale=0.47]{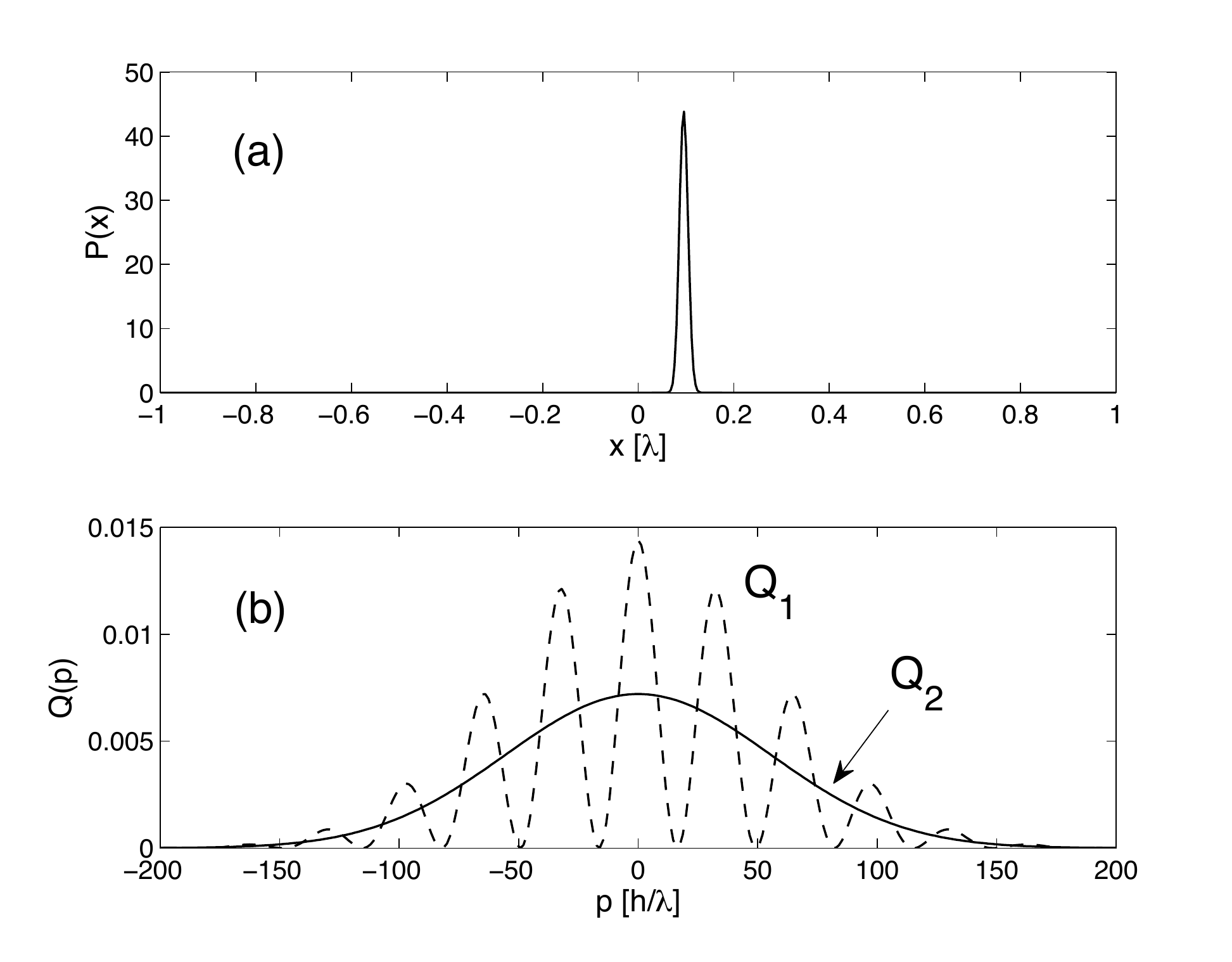}
\caption{The same as in Figure~\ref{fig:localise} but with the particles initially localised before the photons are scattered. (a) There is now only one peak in the relative position or, strictly, an equally weighted mixture of the two peaks, both of which give the same relative momentum distribution. (b) The corresponding relative momentum probability density is shown as a solid line (labelled $Q_2$) and is compared to the result in Figure~\ref{fig:localise} shown as a dashed line (labelled $Q_1$).}
\label{fig:prelocalised}
\end{figure}

The distribution shown in Fig.~\ref{fig:localise}(a) after 150 photons have been detected has two peaks that are symmetric about the origin. This is what we would expect since we have an equal superposition of either particle being to the ``left". If the two particles had a well-defined relative position to begin with, then there would be only one peak in this distribution as shown, for example, in Fig.~\ref{fig:prelocalised}(a). Strictly, in this case we have a mixture of the two different relative positions. This mixture reflects the classical uncertainty in our knowledge of the relative position of the particles based on detecting the scattered photons. The case where the particles are initially delocalised is quite different and gives a coherent superposition of the two relative positions.

The relative-position localisation  process is analogous to the build-up of relative phase between two number state Bose-Einstein condensates when interference patterns are detected between them \cite{Javanainen, Dalibard1997a, Dunningham1999a}. Just like in that case we cannot distinguish from the detected photons whether the position (or phase of the BECs) was well-defined to begin with or created by the measurements. We need a way of doing this in order to experimentally verify that  the localisation process takes place. Although the distinction between Fig.~\ref{fig:localise}(a)  and Fig.~\ref{fig:prelocalised}(a) is clear, an experimentalist would not have direct access to this since the detected photons cannot tell these distributions apart. One possible solution is to look in the conjugate space -- in this case relative momentum. A similar idea has been applied to BECs \cite{Dunningham2006a, Mullin2010a, Mullin2010b}. 

The relative momentum distributions corresponding to the relative position distributions in Figs~\ref{fig:localise}(a) and \ref{fig:prelocalised}(a) are shown as the solid lines in Figs~\ref{fig:localise}(b) and \ref{fig:prelocalised}(b) respectively. For ease of comparison, the result from Fig.~\ref{fig:localise}(b) is superimposed on Fig.~\ref{fig:prelocalised}(b) as a dashed line. We see that the two distributions have the same envelope, but the case where localisation is induced has interference fringes. For particles that  are {\it a priori} perfectly localised, the distribution in Fig.~\ref{fig:prelocalised}(a) would be a delta function and the momentum distribution would be completely flat. We have chosen the relative position distribution shown because it is an upper limit to the width possible based on the photons detected. In other words, it is the `hardest' case to distinguish from that shown in Fig.~\ref{fig:localise}(a). We want to demonstrate that our technique works even in this worst-case scenario.

The measurement scheme is then quite straightforward. After scattering the photons from the particles, we want to distinguish the two relative momentum distributions shown in Fig.~\ref{fig:prelocalised}(b). To do this, we switch off any trapping potential and allow the particles to move freely. By detecting their positions in the $x$-direction after some time of flight, we can infer the $x$-components of their momenta and hence the relative momentum of the particles in that direction. By repeating the whole process from the beginning  many times, we should be able to build up a probability distribution and so distinguish the two cases. However, the stochastic nature of the process means that the particles localise to a different relative position on each run and so the relative momentum fringes are different each time. If we were to just na\"ively add the results from each run, the fringes would wash out. Instead we can use Bayesian analysis to distinguish the two scenarios.

Suppose on a particular run we detected scattered photons on the screen that meant the relative momentum distribution was either $Q_1(p)$ or $Q_2(p)$ depending on whether or not the scattering process induced relative localisation (see Fig.~\ref{fig:prelocalised}(b)). To begin with, we do not know whether the particles are localised or not so we take our prior probability of them initially being localised, $P_l$ to be the same as the prior probability of them not initially being localised, $P_{nl}$, i.e. $P_l = P_{nl} = 0.5$. Now suppose, upon releasing the particles, we measure their relative momentum to be $p_1$. This gives us some information about which scenario is more likely. In particular, Bayes' theorem tells us that the updated probabilities are $P_{nl} \propto Q_1(p_1)\times 0.5$ and $P_{l} \propto Q_2(p_1)\times 0.5$. Normalising, we get
\begin{eqnarray}
P_{nl} =\frac{Q_1(p_1)}{Q_1(p_1)+Q_2(p_1)}\times 0.5
\end{eqnarray}
and $P_l = 1 - P_{nl}$. We can then iterate this process by using these updated probabilities as the prior probabilities in the next step. By repeating many times we increasingly refine our knowledge of which process is occurring.

\begin{figure}
\centering
\includegraphics[scale=0.42]{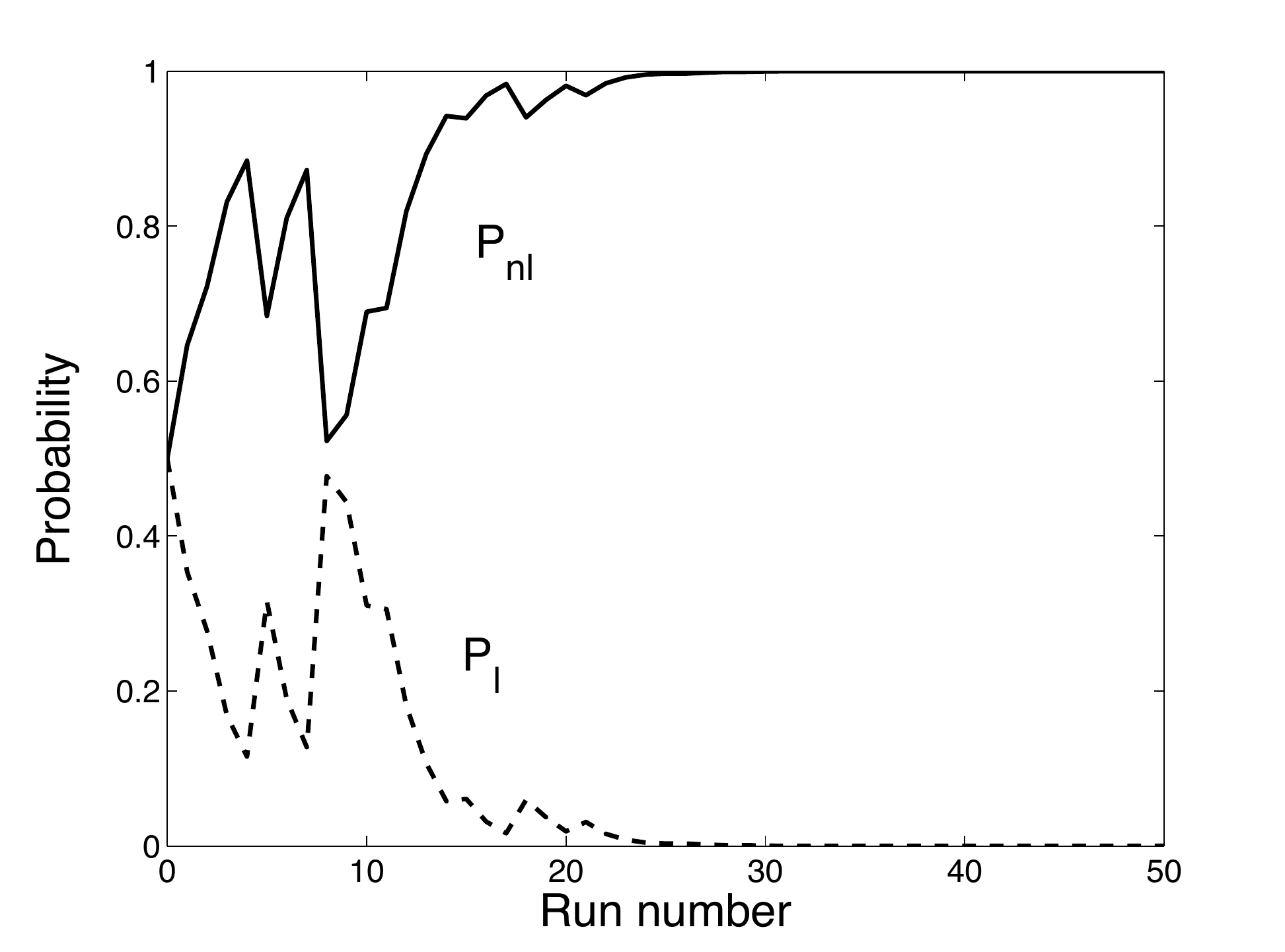}
\caption{A simulated experiment showing the Bayesian analysis of the probability $P_{nl}$, that the photon scattering caused relative localisation of the particles (solid line) and the probability, $P_{l}$, that they were localised to begin with (dashed line). In our simulation, we have taken the particles to start off with no well-defined relative position.}
\label{fig:Bayes}
\end{figure}

A sample simulation is shown in Fig.~\ref{fig:Bayes} for the case that the particles do not initially have a well-defined relative position. We see that initially $P_{nl}=P_l = 0.5$ and that as more and more runs are performed our knowledge of what process is occurring is refined. The probabilities initially jump around for a while before settling down after about 25 runs. The information in this figure is what would be directly accessible to experimentalists and so, in this case, they would be quite certain after about 25 runs that they had observed measurement-induced relative-position localisation.

\begin{figure}
\centering
\includegraphics[scale=0.37]{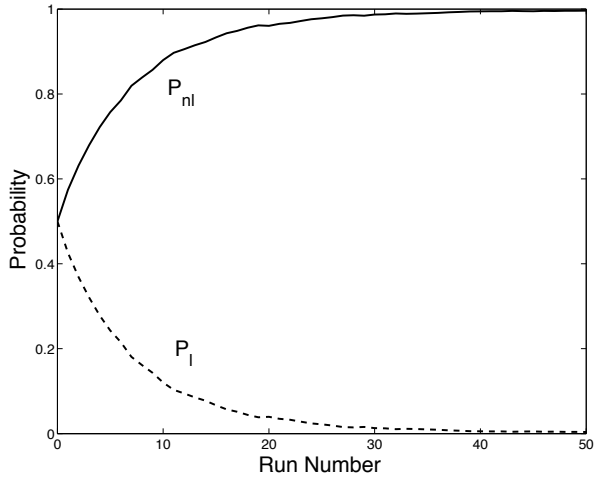}
\caption{As in Figure~\ref{fig:Bayes} but averaged over 300 `experiments' to indicate the average number of runs that would be required to achieve a desired degree of confidence.}
\label{fig:Bayes_average}
\end{figure}

Of course, every experiment would be different due to the stochastic nature of the photon scattering events and the momentum measurements of the particles. So it would be useful to know how many runs on average are likely to be required to achieve a certain degree of confidence. In Fig.~\ref{fig:Bayes_average} we have averaged the results over 300 simulated experiments. We see that the curves are now quite smooth and that after 20 runs we would expect, on average, to be about 95\% confident that measurement-induced relative-position localisation is occurring.

\begin{figure}
\centering
\includegraphics[scale=0.5]{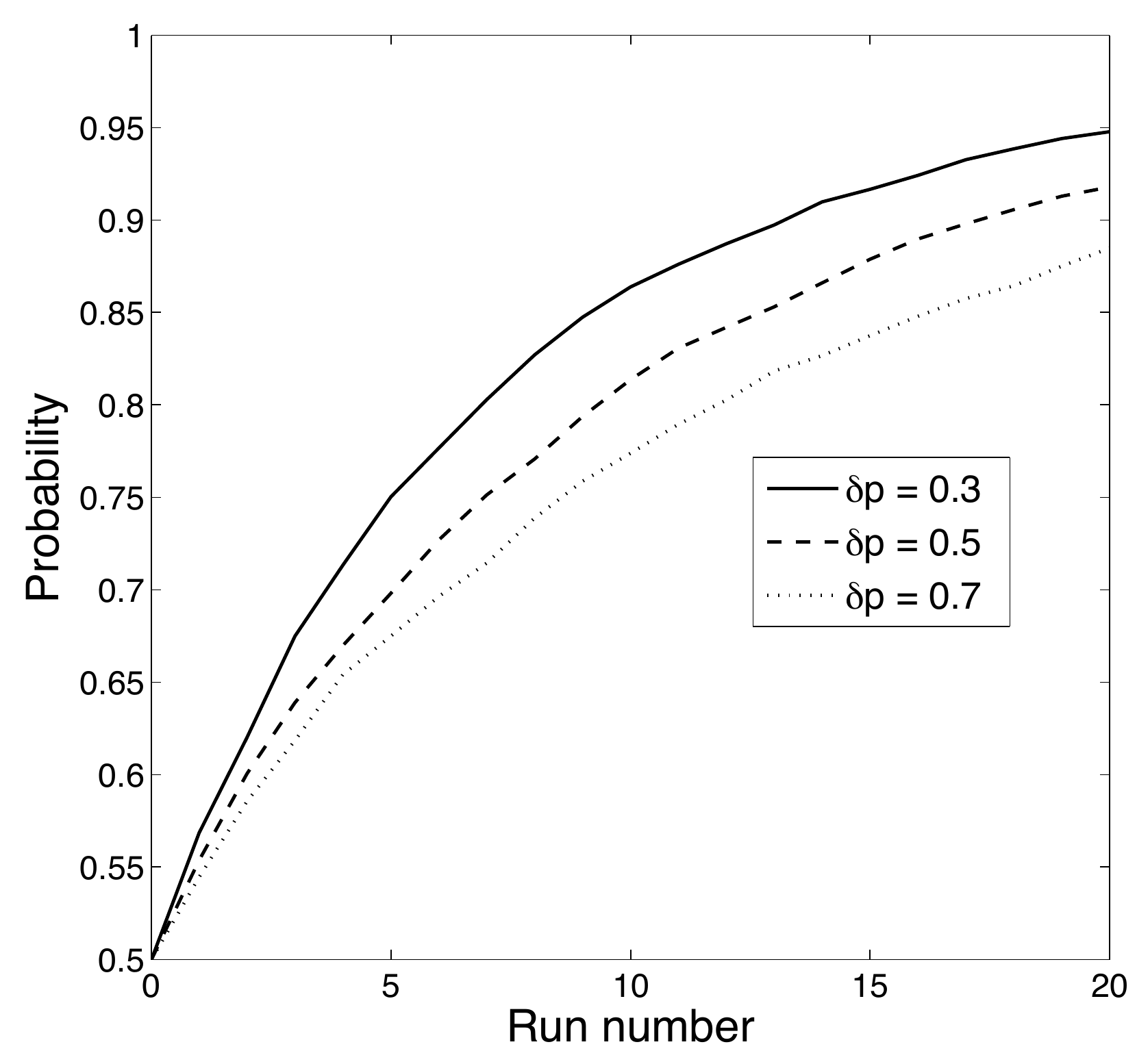}
\caption{This shows $P_{nl}$, the probability that photon scattering caused relative localisation of the particles, as in Figure~\ref{fig:Bayes_average}, but here we also include the effect of imperfect measurements. The three different curves show different values of the precision of the momentum measurement: the values of $\delta p$ here are in units of $[h/\lambda]$. We can see that a resolution of $\delta p=0.5  [h/\lambda]$ is needed for us to be 90\% sure that the two particles were initially delocalised after 20 runs have been completed.}
\label{fig:Different_delta_p3}
\end{figure}

So far we have assumed perfect precision of the momentum measurements that allow us to distinguish the two relative momentum distributions shown in Fig.~\ref{fig:prelocalised}(b). An important question is whether this can still be done when real detectors with imperfect precision are used. To investigate this we convolve the relative momentum probability densities in Fig.~\ref{fig:prelocalised}(b) with Gaussians of different widths, where each width represents a different resolution of the momentum measurement. We then repeat the above analysis using Bayes' theorem, which gives us the results shown in Fig.~\ref{fig:Different_delta_p3}. Not surprisingly, it can be seen that as the measurement precision improves, our knowledge about which process has occurred increases more rapidly with the number of runs. In order to be 90\% sure that the state was initially delocalised after 20 runs, we need to be able to measure the momentum of the particles with a resolution of $\delta p=0.5  [h/\lambda]$ or less.

As discussed above, we propose that the momentum of the particles is measured by switching off any trapping potential and then detecting the particles' positions in the $x$-direction after some time of flight. It is the spatial resolution of this position measurement that we are concerned with, and requiring a momentum resolution of $\delta p=0.5  [h/\lambda]$ translates to a detector spatial resolution requirement of approximately $25 \mu m$ (this assumes we use Rb-87 atoms, illuminated with violet light, which are allowed to fly for a time of $5 ms$ along a detector of length $10 mm$). Using time-of-flight fluorescence imaging it is possible to spatially resolve the position of a single atom with resolution close to $1 \mu m$ \cite{Fuhrmanek2010a}, and furthermore, B\"ucker et. al. achieve single atom detection with efficiency close to unity \cite{Bucker2009a}, so our required momentum measurement should be achievable with current techniques.

\section{Extension to Three Dimensions}

\begin{figure}
\centering
\includegraphics[scale=0.7]{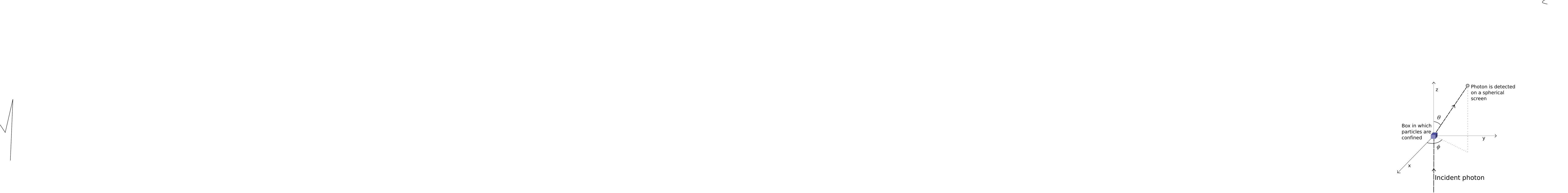}
\caption{This diagram illustrates the experiment in which two massive particles are delocalised over the volume of a cube with side length $d$. Plane-wave light with wavelength $\lambda$ incident along the $z$-axis scatters from the particles and is detected at an angle ($\theta$,$\phi$) on a spherical screen located at a distance $L$ from the particles. For clarity, the diagram is not to scale: we consider the case where $L\gg d$.}
\label{fig:3Dscatter}
\end{figure}

We have reviewed a proof of principle for the localisation of particles caused by entanglement. We now extend this scheme to particles that are allowed to move in three dimensions. Two distinguishable non-interacting particles are initially delocalised within a 3D cube of length $d=\lambda$. The particles are illuminated with plane-wave light of wavelength $\lambda$ incident along the $z$-axis, which scatters from them and is detected at an angle ($\theta$,$\phi$) on a spherical screen located at a distance $L$ from the particles, as shown in Fig.~\ref{fig:3Dscatter}. The initial wave function of the particles is $c(x,y,z)$, where $x$,$y$ and $z$ represent the \emph{relative} position of the two particles in cartesian coordinates. The wavefunction $c(x,y,z)$ is normalised so that  $\int_{}^{}\int_{}^{}\int_{D}^{}|c(x,y,z)|^2dxdydz = 1$, where $D$ represents the box dimensions in which the particles are confined. As in the 1D case, we can neglect the centre-of-mass coordinate of the particles.

We now look at the scattering process: a photon of wavelength $\lambda$ scatters off a particle into angle $(\theta,\phi)$ in spherical coordinates where $\theta$ and $\phi$ are the polar and azimuthal angles, respectively. A deflected photon will impart the following momentum kick on one of the particles:
\begin{gather*}
\Delta p_x = h\sin\theta\cos\phi/\lambda\\
\Delta p_y = h\sin\theta\sin\phi/\lambda\\
\Delta p_z = h(\cos\theta-1)/\lambda
\end{gather*}
where $h$ is Planck's constant. In relative momentum space the particles therefore receive a kick of $\pm \Delta p$ where $\Delta p=(\Delta p_x,\Delta p_y,\Delta p_z)$. Whether they receive a $+\Delta p$ or $-\Delta p$ kick depends on which particle the photon scatters from, but since this cannot be determined we obtain a superposition of both possibilities.

After one scattering event the overall state of the system is given by:

\begin{eqnarray}
\begin{split}
&\Psi(x,y,z,\theta,\phi) = \\
&\begin{cases}
\frac{1}{2\sqrt{\pi}}c(x,y,z) \left( \cos {\frac{2\pi}{\lambda}\Gamma_{x,y,z}(\theta,\phi)} \right)  & \text{if } (\theta,\phi) \not= (0,0) \label{label1} \\ \\
c(x,y,z)A(x,y,z) & \text{if } (\theta,\phi)=(0,0)
\end{cases}
\end{split}
\end{eqnarray}

where:
\begin{eqnarray*}
\Gamma_{x,y,z}(\theta,\phi) =  \left[ x\sin\theta\cos\phi + y\sin\theta\sin\phi + z(\cos\theta - 1) \right].
\label{kicks}
\end{eqnarray*}

The nonscattering coefficient is given by:

\begin{eqnarray*}
A^2(x,y,z) = \frac{1}{4\pi}\int_{0}^{2\pi}\int_{0}^{\pi} \sin\theta'\sin^2 \left( \frac{2\pi}{\lambda}\Gamma_{x,y,z}(\theta',\phi') \right) \, d\theta'd\phi'
\end{eqnarray*}

The localisation process works in the same way as the 1D case. The probability density for detecting a scattered photon at angle $(\theta,\phi)\neq (0,0)$ is $P_S(\theta,\phi)$, whereas for a nonscattered photon the probability density is $P_{NS}(0,0)$:
\begin{align*}
P_S(\theta,\phi) &= \iiint_D |\Psi_{(\theta,\phi) \not= 0}|^2 dxdydz \\
P_{NS} &= \iiint_D |c(x,y,z)|^2 A^2\, dxdydz
\end{align*}
Again this means that detecting a nonscattered photon can actually give us information about the separation of the particles. As before, we generate a random number to see whether the photon is scattered and, if so, at what angle $(\theta_1,\phi_1)$. The (unnormalised) states after the scattering process are as follows, for photons scattered at angle $(\theta_1,\phi_1)$, and non scattered photons, respectively:

\begin{align*}
\Psi_{\theta_1\phi_1}  &= c(x,y,z)\cos \left( \frac{2\pi}{\lambda}\Gamma_{x,y,z}(\theta_1,\phi_1) \right) \\
\Psi_{0 0} &= c(x,y,z)A(x,y,z).
\end{align*}
We then normalise the state and repeat for the next photon.

\begin{figure}
\centering
\includegraphics[scale=0.47]{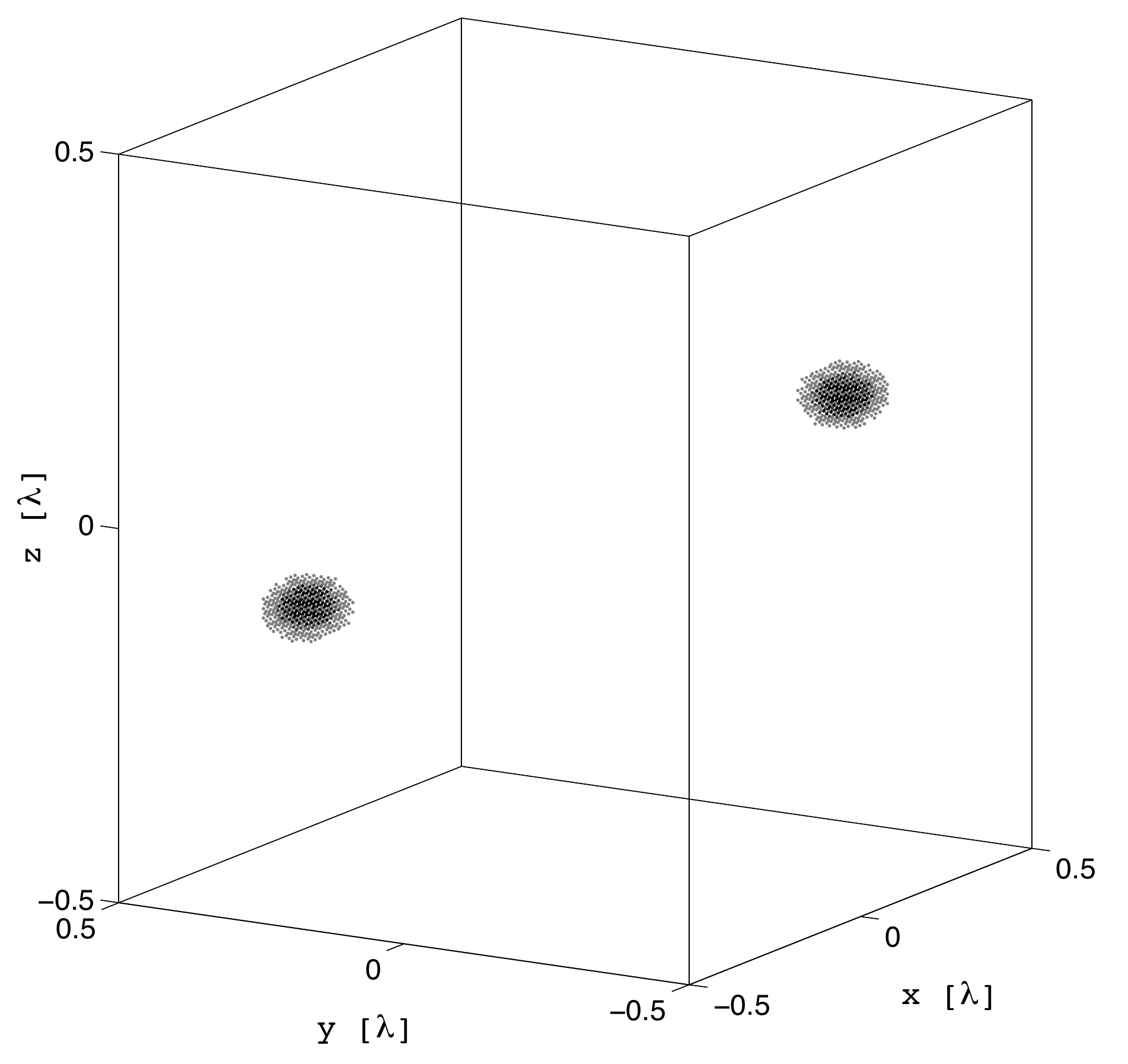}
\caption{This plot shows the probability density $P(x)$, represented by the density and shading of points, for the relative position of the particles after the scattering and detection of 150 photons. The two high density dark clouds show that as in the 1D case, light scattering has caused relative position localisation. The position is given in units of the wavelength, $\lambda$, of the scattered light.}
\label{fig:cloud}
\end{figure}

We have chosen the initial probability density of the relative positions of the particles to be a flat distribution. We find that after successive photons are scattered off the particles, their relative positions localise, as shown by the probability distribution of the two particles in Fig.~\ref{fig:cloud} after 150 photons have been scattered. Again we assume that the 150 photons are all incident on the particles in a sufficiently short time period that we do not need to consider the dynamics of the particles between detection events. The high probability density regions in Fig.~\ref{fig:cloud} are symmetrical about the origin. This reflects the fact that the two particles are interchangeable, and that the localisation is a result of successive superpositions of positive and negative relative momentum kicks. This is the desired result: it shows that scattering induced localisation can be extended to the more realistic case of particles that are allowed to move in three dimension. As in the one dimensional case, it is important to note that the localisation is strictly in $relative$ position space: no $absolute$ position localisation has occurred.

\section{Conclusion and Discussion}

We have demonstrated a simple scheme that should enable experimentalists to unambiguously determine whether scattering events can induce relative position localisation for quantum particles. This is an interesting interpretation for how ambient scattering events could lead to the emergence of classical-like behaviour in quantum systems. We have also extended this scheme from a one dimensional proof of principle to the more real world scenario of three dimensions and considered some practical issues for carrying out the experiment. This idea could have important consequences for our understanding of the boundary between quantum and classical physics and the role of relative observables in nature.\\

This work was partly supported by DSTL (contract number DSTLX1000063869).



\end{document}